\documentclass{ws-procs9x6}

\usepackage{amsmath,graphicx}

\def\bbbr{{\Bbb R}}

\usepackage{amsfonts,amssymb}
\begin{document}

\title{THE MANAKOV SYSTEM AS TWO MOVING INTERACTING CURVES}

\author{N. A. KOSTOV$^{1}$, R. DANDOLOFF$^{2}$, V. S. GERDJIKOV$^{3}$  and G. G. GRAHOVSKI$^{2,3}$}

\address{$^{1}$ Institute of Electronics, Bulgarian Academy
of Sciences, 72 Tsarigradsko chaussee, 1784 Sofia, Bulgaria
\\
$^{2}$ Universit\'e de Cergy-Pontoise, 2 avenue, A. Chauvin,
F-95302, Cergy-Pontoise Cedex, France
\\
$^{3}$Institute for Nuclear Research and Nuclear Energy, Bulgarian
Academy of Sciences, 72 Tsarigradsko chaussee, 1784 Sofia,
Bulgaria}

\begin{abstract}
The two time-dependent Schr\"{o}dinger equations in a potential
$V(s,u)$, $u$ denoting time, can be interpreted geometrically as a
moving interacting curves whose Fermi-Walker phase density is
given by $-(\partial V/\partial s)$. The Manakov model appears as
two moving interacting curves using extended da Rios system and
two Hasimoto transformations.
\end{abstract}


\bodymatter

\section{Introduction}

In recent years, there has been a large interest in the
applications of the Frenet-Serret equations \cite{e60,RogSchief}
for a space curve in various contexts, and interesting connections
between geometry and integrable nonlinear evolution equations have
been revealed. The subject of how space curves evolve in time is
of great interest and has been investigated by many authors.
Hasimoto \cite{h72} showed that the evolution of a thin vortex
filament regarded as a moving space curve can be mapped to the
nonlinear Schr\"{o}dinger equation (NLSE):
\begin{eqnarray}\label{eq:1}
i \Psi_{u}+\Psi_{ss}+\frac{1}{2}|\Psi|^{2}\Psi=0,
\end{eqnarray}
Here, u and s are time and space variables, respectively,
subscripts denote partial derivatives. Lamb \cite{L77} used
Hasimoto transformation to connect other motions of curves to the
mKdV and sine-Gordon equations. Langer and Perline \cite{lr91}
showed that the dynamics of non-stretching vortex filament in
$\bbbr^{3}$ leads to the NLS hierarchy. Motions of curves on
$S^{2}$ and $S^{3}$ were considered by Doliwa and Santini
\cite{DoSa}. Lakshmanan \cite{Laks} interpreted the dynamics of a
nonlinear string of fixed length in $\bbbr^{3}$ through the
consideration of the motion of an arbitrary rigid body along it,
deriving the AKNS spectral problem without spectral parameter.
Recently, Nakayama \cite{Nakaya} showed that the defocusing
nonlinear Schr\"{o}dinger equation, the Regge-Lund equation, a
coupled system of KdV equations and their hyperbolic type arise
from motions of curves on hyperbola in the Minkowski space.
Recently the connection between motion of space or plane curves
and integrable equations has drawn wide interest  and many results
have been obtained \cite{FokGel,dodd,ciel,ChouQu,LanPer,H02,CaIv}.

\section{Preliminaries}\label{sec:2}

\subsection{The Manakov model}\label{sec:2.1}

Time-dependent Schr\"{o}dinger equation in potential $V(s,u)$
\begin{eqnarray}
i\Psi_{u}+\Psi_{ss}-V(s,u)\Psi=0,
\end{eqnarray}
goes into the NLS eq. (\ref{eq:1}) if the potential $V(s,u)=-1/2
|\psi(s,u)|^2$. Similarly, a set of two time-dependent
Schr\"odinger equations:
\begin{eqnarray}
i\Psi_{1,u}+\Psi_{1,ss}-V(s,u)\Psi_{1}=0,\quad
i\Psi_{2,u}+\Psi_{2,ss}-V(s,u)\Psi_{2}=0,
\end{eqnarray}
where $V(s,u)=-|\Psi_{1}|^{2} -|\Psi_{2}|^{2}$, can be viewed as
the Manakov system:
\begin{eqnarray}
&&i\Psi_{1,u}+\Psi_{1,ss}+(|\Psi_{1}|^{2}
+|\Psi_{2}|^{2})\Psi_{1}=0,\label{ManakovSys1}\\
&&i\Psi_{2,u}+\Psi_{2,ss}+(|\Psi_{1}|^{2}+|\Psi_{2}|^{2})\Psi_{2}=0.
\label{ManakovSys2}
\end{eqnarray}
It is convenient to use two Hasimoto transformations \cite{h72}
\begin{eqnarray}
\Psi_{i}=\kappa_{i}(s,u) \exp\left[i\int^{s}\tau_{i}(s',u)d
s'\right],\quad i=1,2,
\end{eqnarray}
in Eqs. (\ref{ManakovSys1}), (\ref{ManakovSys2}). Equating
imaginary and real parts, this leads to the coupled partial
differential equations (extended daRios system \cite{r1906})
\begin{eqnarray}
&&\kappa_{i,u}=-(\kappa_{i}\tau_{i})_{s}-\kappa_{i,s}\tau_{i},\quad
i=1,2 ,\label{daRios1}\\
&&\tau_{i,u}=\left[\frac{\kappa_{i,ss}}{\kappa_{i}}-\tau_{i}^{2}\right]_{s}-V(s,u)_{s},
\label{daRios2}
\end{eqnarray}
where
\begin{eqnarray}
V(s,u)=-|\Psi_{1}|^{2}
-|\Psi_{2}|^{2}=-\kappa_{1}^{2}-\kappa_{2}^2.
\end{eqnarray}

\subsection{Soliton curves}\label{sec:2.2}

A three dimensional space curve is described in parametric form by
a position vectors ${\bf r}_{i}(s), i=1,2$, where s is the
arclength. Let ${\bf t}_{i}=(\partial {\bf r}_{i}/\partial s),
i=1,2$ be the two unit tangent vectors along the two curves. At a
given instant of time the triads of unit orthonormal vectors
$({\bf t}_{i},{\bf n}_{i},{\bf b}_{i})$, where ${\bf n}_{i}$ and
${\bf b}_{i}$ denote the normals and binormals, respectively,
satisfy the  Frenet-Serret equations for two curves
\begin{eqnarray} \label{FSeq}
{\bf t}_{i,s}=\kappa_{i} {\bf n_{i}},\quad {\bf
n}_{i,s}=-\kappa_{i} {\bf t}_{i}+ \tau_{i}{\bf b}_{i},\quad {\bf
b}_{i,s}=-\tau_{i} {\bf n}_{i},\quad i=1,2 ,
\end{eqnarray}
$\kappa_{i}$ and $\tau_{i}$ denote, respectively the two
curvatures and torsions of the curves. A moving curves are
described by $r_{i}(s,u)$, where u denote time. The temporal
evolution of two triads corresponding to a given value $s$ can be
written in the general form as
\begin{eqnarray}
{\bf t}_{i,u}=g_{i} {\bf n}_{i}+h_{i} {\bf b}_{i},\quad {\bf
n}_{i,u}=-g_{i} {\bf t}_{i} + \tau^{0}_{i}{\bf b}_{i},\quad {\bf
b}_{i,u}=-h_{i} {\bf t}_{i} - \tau^{0}_{i}{\bf n}_{i},
\end{eqnarray}
where the coefficients $g_{i}$,$h_{i}$ and $\tau_{i}^{0}$, as well
as $\kappa_{i}$ and $\tau_{i}$, are functions of $s$ and $u$.
\begin{eqnarray}
\left(%
\begin{array}{c}
  {\bf t}_{i} \\
  {\bf n}_{i} \\
  {\bf b}_{i} \\
\end{array}%
\right)_{s}
=\left(%
\begin{array}{ccc}
  0 & \kappa_{i} & 0 \\
  -\kappa_{i} & 0 & \tau_{i} \\
  0 & -\tau_{i} & 0 \\
\end{array}%
\right) \left(%
\begin{array}{c}
  {\bf t}_{i} \\
  {\bf n}_{i} \\
  {\bf b}_{i} \\
\end{array}%
\right),\quad
\left(%
\begin{array}{c}
  {\bf t}_{i} \\
  {\bf n}_{i} \\
  {\bf b}_{i} \\
\end{array}%
\right)_{u}
=\left(%
\begin{array}{ccc}
  0 & g_{i} & h_{i} \\
  -g_{i} & 0 & \tau^{0}_{i} \\
  -h_{i} & -\tau^{0}_{i} & 0 \\
\end{array}%
\right) \left(%
\begin{array}{c}
  {\bf t}_{i} \\
  {\bf n}_{i} \\
  {\bf b}_{i} \\
\end{array}%
\right).\nonumber
\end{eqnarray}
Introduce
\begin{eqnarray}
L_{i} =\left(%
\begin{array}{ccc}
  0 & \kappa_{i} & 0 \\
  -\kappa_{i} & 0 & \tau_{i} \\
  0 & -\tau_{i} & 0 \\
\end{array}%
\right),\qquad M_{i}
=\left(%
\begin{array}{ccc}
  0 & g_{i} & h_{i} \\
  -g_{i} & 0 & \tau^{0}_{i} \\
  -h_{i} & -\tau^{0}_{i} & 0 \\
\end{array}%
\right)
\end{eqnarray}
and $\Delta {\bf t}_{i} \equiv ({\bf t}_{i,su}-{\bf t}_{i,us})$, $
\Delta {\bf n}_{i} \equiv ({\bf n}_{i,su}-{\bf n}_{i,us})$, and
$\Delta {\bf b}_{i} \equiv ({\bf b}_{i,su}-{\bf b}_{i,us}) $. Then
\begin{eqnarray}
\left(%
\begin{array}{c}
 \Delta {\bf t}_{i}  \\
 \Delta {\bf n}_{i}  \\
 \Delta {\bf b}_{i}  \\
\end{array}%
\right) =&&\left( \partial_{s} M_{i}-\partial_{u}
L_{i}+[L_{i},M_{i}] \right)
\left(%
\begin{array}{c}
  {\bf t}_{i} \\
  {\bf n}_{i} \\
  {\bf b}_{i} \\
\end{array}%
\right)\nonumber\\&& =
\left(%
\begin{array}{ccc}
  0 & \alpha^{1}_{i} & \alpha^{2}_{i} \\
 -\alpha^{1}_{i} & 0 & \alpha^{3}_{i} \\
- \alpha^{2}_{i} & -\alpha^{3}_{i} & 0 \\
\end{array}%
\right) \left(%
\begin{array}{c}
  {\bf t}_{i} \\
  {\bf n}_{i} \\
  {\bf b}_{i} \\
\end{array}%
\right),
\end{eqnarray}
where
\begin{eqnarray}
\alpha^{1}_{i}=\kappa_{i,u}g_{i,s}+h_{i}\tau_{i},\,
\alpha^{2}_{i}=-h_{i,s}+\kappa_{i}\tau_{i}^{0}-g_{i}\tau_{i},\,
\alpha^{3}_{i}=\tau_{i,u}-\tau_{i,s}-\kappa_{i} h_{i}.
\end{eqnarray}
\begin{eqnarray}\label{KappaTau}
\kappa_{i,u}=g_{i,s}-h_{i}\tau_{i},\qquad
\tau_{i}^{0}=(h_{i,s}+g_{i}\tau_{i})/\kappa_{i},
\end{eqnarray}
A generic curve evolution must satisfy the geometric constraints
\begin{eqnarray}
 \Delta {\bf t}_{i}\cdot( \Delta {\bf n}_{i}\times  \Delta {\bf
b}_{i})=0, \label{constraint}
\end{eqnarray}
i.e.,  $\Delta {\bf t}_{i}$, $ \Delta {\bf n}_{i}$ and $ \Delta
{\bf b}_{i}$ must remain coplanar vectors under time involution.
Further, since Eq. (\ref{constraint}) is automatically satisfied
for $\Delta {\bf t}_{i}=0$, we see that $\Delta {\bf n}_{i}$ and
$\Delta {\bf b}_{i}$ need not necessarily vanish. In addition, we
see from (\ref{constraint}) that $\Delta {\bf t}_{i}=0$ implies
$\alpha_{i}^{1}=\alpha_{i}^{2}=0$, so that
\begin{eqnarray}
\Delta {\bf n}_{i}=\alpha_{i}^{3} \Delta {\bf b}_{i},\quad \Delta
{\bf b}_{i}=\alpha_{i}^{3} \Delta {\bf n}_{i}\quad
g_{i}=-\kappa_{i} \tau_{i},\qquad h_{i}=\kappa_{i,s}.
\end{eqnarray}
Substituting these in the second equation in (\ref{KappaTau})
gives
\begin{eqnarray}
\tau_{i}^{0}=\left[\frac{\kappa_{i,ss}}{\kappa_{i}}-\tau_{i}^{2}\right],
\end{eqnarray}
hence Eq. (\ref{daRios1}) yields
$(\tau_{i,u}-\tau^{0}_{i,s})=-V(s,u)_{s}=(\kappa_{1}^{2}+\kappa_{2}^2)_{s}$.
Next  there is an underlying angle anholonomy \cite{bd99,bbd90} or
'Fermi-Walker phase' $\delta\Phi^{FW}=(\tau_{i,u}-\tau^{0}_{i,s})
dsdu$ with respect to its original orientation, when $s$ and $u$
change along an infinitesimal closed path of area $dsdu$.

\section{Integration of the extended da Rios system}
The coupled nonlinear equations (\ref{daRios1}),(\ref{daRios2})
constitute the extended Da Rios system as derived in \cite{r1906}
. The solutions of (\ref{daRios1}),(\ref{daRios2}) with
$\kappa(\xi)$ and $\tau=\tau(\xi)$, where $\xi=s-c\,u$ are simple.
On substitution, we obtain
\begin{eqnarray}
&&c\,\kappa_{i,\xi}=2\kappa_{i,\xi}\tau_{i}+\kappa_{i}\tau_{i,\xi},\quad
\xi=s-cu, i=1,2 , \\&& -c \tau_{i,\xi}=\left[
-\tau_{i}^2+\frac{\kappa_{i,\xi\xi}}{\kappa_{i}}+\kappa_{1}^{2}+\kappa_{2}^{2}
\right]_{\xi}, \qquad\tau_{i}=\frac{c}{2}.
\end{eqnarray}
where we use the boundary condition $\kappa_{i}\rightarrow 0, i=1,2$
as $s\rightarrow \infty$. Hence $\kappa_{i}$ obey the nonlinear
oscillator equations
\begin{eqnarray}
\kappa_{i,\xi\xi}+
\left(\sum_{j=1}^{2}\kappa_{j}^{2}\right)\kappa_{i}=a_{i}\kappa_{i},\quad
i=1,2. \label{oscillator7}
\end{eqnarray}
where $a_{i}, i=1,2$ are arbitrary constants.

{\bf Example 1} One soliton solutions of the Manakov system are
given by
\begin{eqnarray}
&&\Psi_{1}=\sqrt{2a}\,\epsilon_{1}\mbox{e}^{i(\frac{1}{2}
c(s-s_{0})+(a-\frac{1}{4} c^2) u)}\mbox{sech}(\sqrt{a}(s-s_{0}-ct))\\
&&\Psi_{2}=\sqrt{2a}\,\epsilon_{2}\mbox{e}^{i(\frac{1}{2}
c(s-s_{0})+(a-\frac{1}{4} c^2) u)}\mbox{sech}(\sqrt{a}(s-s_{0}-ct)),
\end{eqnarray}
and $|\epsilon_{1}|^{2}+|\epsilon_{2}|^{2}=1$

We first note that for Manakov system, the expressions for the
curvatures $\kappa_{i}, i=1,2$ and the torsions $\tau_{i}, i=1,2$
for the moving curves corresponding to a one soliton solutions of
the Manakov system are given by
\begin{eqnarray}
\kappa^2=\kappa_{1}^{2}+\kappa_{2}^{2}=\sqrt{2a}\,\mbox{sech}(\sqrt{a}(s-s_{0}-ct)),\quad
\tau_{1}=\tau_{2}=\frac{1}{2}c.
\end{eqnarray}
and
\begin{eqnarray}
\kappa_{1}=\sqrt{2a}\,\epsilon_{1}\mbox{sech}(\sqrt{a}(s-s_{0}-ct)),\quad
\kappa_{2}=\sqrt{2a}\,\epsilon_{2}\mbox{sech}(\sqrt{a}(s-s_{0}-ct)).\nonumber
\end{eqnarray}

\begin{figure}
\begin{center}
\includegraphics[width=10cm,height=8cm]{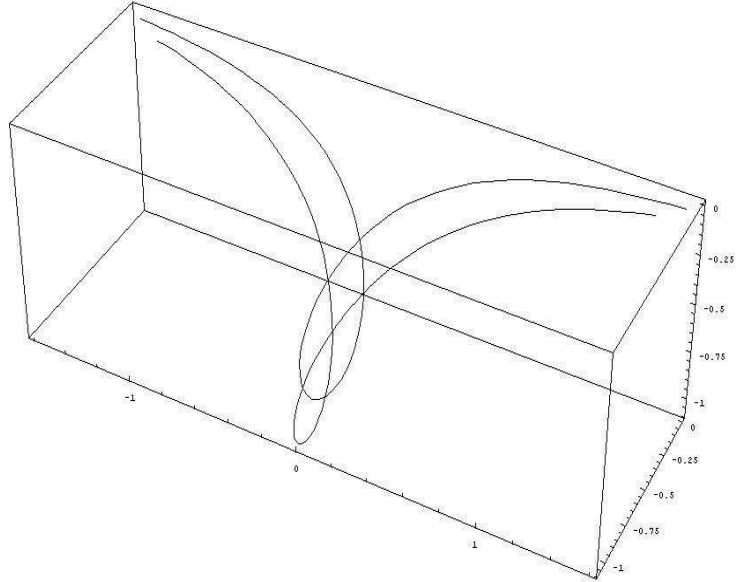}
\caption{Two curves (\ref{oneSolManSys}) of one soliton solution of
Manakov system, $\epsilon_{1}=\sqrt{2}/\sqrt{3}$,
$\epsilon_{2}=1/\sqrt{3}$ } \label{fig1}
\end{center}
\end{figure}

{\bf Example 2} One special solution of Manakov system is written
by
\begin{eqnarray}
\kappa_{1}=C_{1}\mbox{cn}(\alpha \xi,k),\qquad
\kappa_{2}=C_{2}\mbox{cn}(\alpha \xi,k),
\end{eqnarray}
where
\begin{eqnarray}
\alpha^2=\frac{a_{1}}{2k^2-1},\quad C_{1}^{2}+C_{2}^{2}= 2\alpha^2
k^2, \quad a_{1}=a_{2}=a,
\end{eqnarray}
In the limit $k\rightarrow 1$ we obtain the well known {\it
Manakov} soliton solution
\begin{eqnarray}
&&\Psi_{1}=\frac{ \sqrt{2a} \epsilon_{1}
\exp\left\{i\left(\frac{1}{2}c(s-s_{0})+(a-\frac{1}{4}
c^{2})u\right) \right\} }
{\mbox{ch}(\sqrt{a} (s-s_{0}-ct)) },\nonumber \\
&&\Psi_{2}=\frac{ \sqrt{2a} \epsilon_{2}
\exp\left\{i\left(\frac{1}{2}c(s-s_{0})+(a-\frac{1}{4}
c^{2})u\right) \right\} } {\mbox{ch}(\sqrt{a} (s-s_{0}-ct)) }.
\nonumber
\end{eqnarray}
Here we introduce the following notations
\begin{eqnarray}
|\epsilon_{1}|^2+|\epsilon_{2}|^2=1, \quad \zeta_{1}=\frac{1}{2}
c+i\sqrt{a} =\xi_{1}+i\eta_{1},
\end{eqnarray}
where $s_{0}$ is the position of soliton,
$(\epsilon_{1},\epsilon_{2})$ are the components of polarization
vector. The real part of $\zeta_{1}$ i.e. $c/2$ gives us the
soliton velocity while the imaginary part of $\zeta_{1}$, i.e.
$\sqrt{2a}$, gives the soliton amplitude and width.

{\bf Example 3} Integrating (\ref{FSeq}) for two unit tangent
vectors along the  curves ${\bf t}_{i}=(\partial {\bf
r}_{i}/\partial s), i=1,2$  for position vectors ${\bf r}_{i}(s),
i=1,2$ we obtain
\begin{eqnarray}
{\bf r}_{j}=\left(\begin{array}{c}
  \frac{s}{2}-\frac{\epsilon_{j}}{\epsilon_{j}^{2}+\frac{1}{2} c} \tanh{\left(\epsilon_{j}(s-cu)\right)} \\
  -\frac{\epsilon_{j}}{\epsilon_{j}^{2}+\frac{1}{2} c} \mbox{sech}{\left(\epsilon_{j}(x-cu)\right)}
  \cos{\left(\frac{1}{2}cs+(\epsilon_{j}^{2}-\frac{1}{4} c^2)u \right)} \\
   -\frac{\epsilon_{j}}{\epsilon_{j}^{2}+\frac{1}{2} c} \mbox{sech}{\left(\epsilon_{j}(x-cu)\right)}
   \sin{\left(\frac{1}{2}cs+(\epsilon_{j}^{2}-\frac{1}{4} c^2)u \right)} \\
\end{array}\right),\quad j=1,2 , \label{oneSolManSys}
\end{eqnarray}
and $\epsilon_{1}=\cos{\alpha},\,\epsilon_{2}=\sin{\alpha}$, where
$\alpha$ is arbitrary positive number.

 {\bf Example 4} Let
$u(x)=6\wp(\xi+\omega^{\prime})$ be the two-gap Lam\'e potential
with simple periodic spectrum (see for example \cite{ek94})
\begin{equation}
\lambda_{0}=-\sqrt{3g_{2}},\quad \lambda_{1}=-3e_{0}, \quad
\lambda_{2}=-3e_{1}, \quad  \lambda_{3}=-3e_{2}, \quad
\lambda_{4}=\sqrt{ 3g_{2}}.
\end{equation}
and the corresponding Hermite polynomial have the form
\begin{equation}
F(\wp(\xi+\omega^{\prime}),\lambda)=\lambda^{2}-
3\wp(\xi+\omega^{\prime})\lambda+
9\wp^{2}(\xi+\omega^{\prime})-\frac{9}{4}g_{2} .  \label{HerPol}
\end{equation}
Consider the genus $2$ nonlinear anisotropic oscillator
(\ref{oscillator7}) with Hamiltonian
\begin{equation}
H=\frac{1}{2}(p_{1}^{2}+p_{2}^{2})+\frac{1}{4}(\kappa_{1}^{2}+\kappa_{2}^{2})^{2}-
\frac{1}{2}(a_{1}\kappa_{1}^{2}+a_{2}\kappa_{2}^{2}),
\end{equation}
where $(\kappa_{i},p_{i})$, $i=1,2$ are canonical variables with
$p_{i}=\kappa_{i,x}$ and $a_{1},a_{2}$ are arbitrary constants.
The simple solutions of these system are given in terms of Hermite
polynomial
\begin{equation}
\kappa_1^2=2\frac{F(x,\tilde{\lambda}_{1})} {\tilde{\lambda}_{2}-\tilde{\lambda}%
_{1}} ,\quad \kappa_2^2=2\frac{F(x,\tilde{\lambda}_{2})} {\tilde{\lambda}_{1}-%
\tilde{\lambda}_{2}} ,
\end{equation}
Let us list the corresponding solutions

{\bf (A)} Periodic solutions in terms of single Jacobian elliptic
function

The nonlinear anisotropic oscillator admits the following
solutions:
\begin{eqnarray}
\kappa_1  =  C_1 \mbox{sn}(\alpha \xi, k), \qquad \kappa_2  = C_2
\mbox{cn}(\alpha \xi, k).  \label{onegap}
\end{eqnarray}
Here the amplitudes $C_1$, $C_2$ and the temporal pulse-width
$1/\alpha$ are defined by the parameters $a_{1}$ and $a_{2}$ as
follows:
\begin{eqnarray}
\alpha^2 k^2  =  a_{2}-a_{1} , \quad C^{2}_{1}  =  a_{2} +
\alpha^2 - 2\alpha^2 k^2 , \quad C^{2}_{2} = a_{1} +
\alpha^{2}+\alpha^2 k^2 ,
\end{eqnarray}
where $0 < k < 1$.

Following our spectral method it is clear, that the solutions
(\ref{onegap}) are associated with eigenvalues $\lambda_2 = - e_2$
and $\lambda_3 = - e_3$ of one -- gap Lam\'e potential.

{\bf (B)} Periodic solutions in terms of products of Jacobian
elliptic functions

Another solution is defined by \cite{ft89}
\begin{eqnarray}
\kappa_1  =  C \mbox{dn}(\alpha \xi, k) \mbox{sn}(\alpha \xi, k),
\qquad \kappa_2  =  C \mbox{dn}(\alpha \xi ,k ) \mbox{cn}(\alpha
\xi, k), \label{Flor}
\end{eqnarray}
where $\mbox{sn}$,$\mbox{cn}$, $\mbox{dn}$ are the standard
Jacobian elliptic functions, $k$ is the modulus of the elliptic
functions $ 0 < k < 1$. The wave characteristic parameters:
amplitude $C$, temporal pulse-width $1/\alpha$ and $k$ are related
to the physical parameters and, $k$ through the following
dispersion relations
\begin{eqnarray}
C^{2}  =  \frac{2}{5} (4a_{2} - a_{1}) , \quad \alpha^{2}  =
\frac{1}{15}(4a_{2}-a_{1}) , \quad k^{2}  =  \frac {5
(a_{2}-a_{1})} {4a_{2}-a_{1}} .
\end{eqnarray}
We have found the following solutions of the nonlinear oscillator
\cite{ku92}
\begin{eqnarray}
\kappa_1  =  C\alpha^2 k^2 \mbox{cn}(\alpha \xi,k)\mbox{sn}(\alpha
\xi,k), \quad \kappa_2  =  C\alpha^2 \mbox{dn}^2 (\alpha \xi, k) +
C_{1} \label{UzKos}
\end{eqnarray}
where $C$, $C_1$, $\alpha$ and $k$ are expressed through
parameters $a_{1}$ and $a_{2}$ by the following relations
\begin{eqnarray}
C^2 & = & \frac {18} {a_{2}-a_{1}}, \quad \alpha^2  =  \frac
{1}{10} \left( 2 a_{2}-3a_{1}+
\sqrt{\frac{5}{3}(a_{2}^{2}-a_{1}^2) } \right)
 \nonumber \\
k^2 & = & \frac {2 \sqrt { \frac{5}{3} (a_{2}^{2}-a_{1}^{2})}}
{\sqrt{\frac{5}{3} (a_{2}^{2}-a_{1}^{2})}+2a_{2}-3a_{1}},\quad C_1
=  \frac {C}{30} (4a_{1}-a_{2}),.
\end{eqnarray}

{\bf (C)} Periodic solutions associated with the two-gap
Treibich-Verdier potentials. Below we construct the two periodic
solutions associated with the Treibich-Verdier potential. Let us
consider the potential
\begin{equation}
u(x)=6\wp(\xi+\omega^{\prime})+2{\frac{(e_1-e_2)(e_1-e_3)}{\wp(\xi+\omega^{%
\prime})-e_1}},  \label{tv4}
\end{equation}
and construct the solution in terms of Lam\'e polynomials
associated with the eigenvalues
$\tilde{\lambda}_1,\tilde{\lambda}_2$, $\tilde{\lambda}_1 >
\tilde{\lambda}_2$
\begin{eqnarray}
\tilde{\lambda}_1=e_2+2e_1+2\sqrt{(e_1-e_2)(7e_1+2e_2)}, \\
\tilde{\lambda}_2=e_3+2e_1+2\sqrt{(e_1-e_3)(7e_1+2e_3)}. \nonumber
\label{zz}
\end{eqnarray}
The finite and real solutions $q_1,q_2$ have the form
\begin{eqnarray}
\kappa_1= C_{1}\mbox{sn}(\xi,k)\mbox{dn}(\xi,k)
+C_{2}\mbox{sd}(\xi,k) ,\, \kappa_2=
C_{3}\mbox{cn}(\xi,k)\mbox{dn}(\xi,k) +C_{4}\mbox{cd}(z,k),
\nonumber
\end{eqnarray}
where $C_{i}$, $i=1,\ldots 4$ are constants and have important
geometrical interpretation \cite{ek94} and  $\mbox{sd}$,
$\mbox{cd}$, are standard Jacobian elliptic functions.  The
concrete expressions in terms of $k,\tilde{
\lambda}_{1},\tilde{\lambda_{2}}$ are given in
\cite{ceek95,chr:eil:eno:kos}

In a similar way we can find the elliptic solution associated with
the eigenvalues
\begin{eqnarray}
\tilde{\lambda}_1&=&e_2+2e_1+2\sqrt{(e_1-e_2)(7e_1+2e_2)},\quad \tilde{%
\lambda}_2=-6e_1,  \label{zz2}
\end{eqnarray}
We have
\begin{eqnarray}
\kappa_1=\tilde{C}_{1}\mbox{dn}^{2}(\xi,k),\qquad \kappa_2=
C_{1}\mbox{sn}(\xi,k)\mbox{dn}(\xi,k) +C_{2}\mbox{sd}(\xi,k) ,
\label{ee3}
\end{eqnarray}
where $\tilde{C}_{1}, C_{1}, C_{2}$ are given in \cite{ceek95,
chr:eil:eno:kos}.

The general formula for elliptic solutions of genus $2$ nonlinear
anisotropic oscillator is given by \cite{ceek95}
\begin{eqnarray}
\kappa_1^2&=&{\frac{1}{\tilde{\lambda}_2-\tilde{\lambda}_1}}
\left(2\tilde{\lambda}_1^2+2\tilde{\lambda}_1\sum_{i=1}^N
\wp(\xi-x_i) \right. \nonumber
\\ &&+\left.6\sum_{1\leq i< j\leq N}\wp(\xi-x_i)\wp(\xi-x_j)-{\frac{Ng_2}{4}}+
\sum_{1\leq i< j\leq 5}\lambda_i\lambda_j\right),  \nonumber \\
\kappa_2^2&=&{\frac{1}{\tilde{\lambda}_1-\tilde{\lambda}_2}}
\left(2\tilde{\lambda}_2^2+2\tilde{\lambda}_2\sum_{i=1}^N
\wp(\xi-x_i) \right. \nonumber
\\ &&+\left.6\sum_{1\leq i< j\leq N}\wp(\xi-x_i)\wp(\xi-x_j)-{\frac{Ng_2}{4}}+
\sum_{1\leq i< j\leq 5}\lambda_i\lambda_j\right),  \nonumber
\end{eqnarray}
where $x_{i}$ are solutions of equations $\sum_{i\neq
j}\wp'(x_{i}-x_{j})=0, j=1,\ldots, N$.

\section{Extended da Rios-Betchov system}
Following Betchov we can derive the system of equations, which may
be reduced to those for a two fictitious gases with negative
pressures accompanied with two complicated nonlinear dispersive
stresses. Introducing four new variables
$\rho_{1}=\kappa_{1}^{2}$, $\rho_{1}=\kappa_{1}^{2}$,
$u_{1}=2\tau_{1}$, $u_{2}=2 \tau_{2}$  using extended Da Rios
system (\ref{daRios1}), (\ref{daRios2}) we obtain
\begin{eqnarray} \label{Betchov}
&&\frac{\partial\rho_{1}}{\partial u}+\frac{\partial
(\rho_{1}u_{1})}{\partial s}=0,\qquad
\frac{\partial\rho_{2}}{\partial
u}+\frac{\partial(\rho_{2}u_{2})}{\partial s}=0, \nonumber \\&&
\frac{\partial(\rho_{1} u_{1})}{\partial u}+\frac{\partial
}{\partial s}\left[\rho_{1}
u_{1}^2-(\rho_{1}^2+\rho_{2}^2)-\rho_{1}\frac{\partial^{2}}{\partial
s^2}(\log \rho_{1})\right]=0,\nonumber \\&&
\frac{\partial(\rho_{2} u_{2})}{\partial u}+\frac{\partial
}{\partial s}\left[\rho_{2}
u_{2}^2-(\rho_{1}^2+\rho_{2}^2)-\rho_{2}\frac{\partial^{2}}{\partial
s^2}(\log \rho_{2})\right]=0. \nonumber
\end{eqnarray}

\section{HF system is gauge equivalent to Manakov system}

The vector nonlinear Schr\"{o}dinger equation is associated with
type ${\bf A.III}$ symmetric space ${\rm SU(n+1)}/{\rm
S(U(1)}\otimes {\rm U(n)})$. The special case $n=2$ of such
symmetric space is associated with the famous Manakov system
\cite{ma74}.

Let us first fix the notations and the normalizations of the basis
of ${\frak  g} $. By $\Delta _+ $ ($\Delta _- $) we shall denote
the set of positive (negative) roots of the algebra with respect
to some ordering in the root space. By $\{E_{\alpha }, H_i\} $,
$\alpha \in \Delta $, $i=1 \dots r $ we denote the Cartan--Weyl
basis of ${\frak  g} $ with the standard commutation relations
\cite{Helg}. Here $H_i$ are Cartan generators dual to the basis
vectors $e_i$ in the root space.  The root system is invariant
under the action of the Weyl group ${\frak W}({\frak g}) $ of the
simple Lie algebra ${\frak g} $ \cite{Helg}.

Let us now consider the gauge equivalent systems. The notion of
gauge equivalence allows us to associate with the vector nonlinear
Schr\"{o}dinger equation an equivalent equation solvable by the
ISM for the gauge equivalent linear problem \cite{ForKu*83}:
\begin{eqnarray}\label{eq:2.3}
\tilde{L}\tilde{\psi }(x,t,\lambda )= \left(i{d  \over dx}-\lambda
{\cal  S}(x,t) \right) \tilde{\psi }(x,t,\lambda
)=0, \nonumber\\
\tilde{M}\tilde{\psi }(x,t,\lambda )= \left(i{d  \over
dt}-\lambda^2 {\cal S}-\lambda {\cal S}_{x} {\cal S}(x,t) \right)
\tilde{\psi }(x,t,\lambda )=0,
\end{eqnarray}
where
\begin{eqnarray}\label{eq:2.4}
&&\tilde{\psi }(x,t,\lambda ) = \psi _0^{-1}\psi (x,t,\lambda ),
\quad{\cal
S}(x,t)=\sum_{\alpha=1}^{r}(S_{\alpha}E_{\alpha}+S_{\alpha}^{*}E_{-\alpha})
+\sum_{j=1}^{r} S_{j} H_{j} ,\nonumber\\&& {\cal  S}(x,t) =
\mbox{Ad}_{\hat{\psi }_0} J \equiv \psi _0^{-1}J\psi _0(x,t),
\qquad J= \sum_{s=1}^n H_s,
\end{eqnarray}
and $\psi _0=\psi (x,t,0) $ is the Jost solution at $\lambda =0 $.
The zero-curvature condition $[\tilde{L},\tilde{M}]=0 $ is
equivalent to $i {\cal  S}_t-[{\cal S},{\cal S}_{xx}]=0.$ with
${\cal S}^2=I_{n}$.

\section{Conclusions}

In this paper the Manakov model is interpreted as two moving
interacting curves. We derive new extended Da Rios system and
obtain the soliton, one-, and two-phase periodic solution of two
thin vortex filaments in an incompressible inviscid fluid. The
solution was explicitly given in terms of Weierstrass and Jacobian
elliptic functions.

\section*{Acknowledgements} The present work is supported by the
National Science Foundation of Bulgaria,  contract No F-1410. The
work of one of us GGG is supported also by the Bulgarian National
Scientific Foundation Young Scientists scholarship for the project
"Solitons, differential geometry and biophysical models".

\end{document}